\documentclass[prc,aps,floats,floatfix,twocolumn,showpacs,nofootinbib,%
superscriptaddress]{revtex4}
\usepackage{amsmath,amssymb,amsfonts}
\usepackage{graphicx}% Include figure files
\usepackage{hhline}
\usepackage{bm}
\usepackage{amsmath}
\usepackage{epsfig}

\newcommand{\gvec}[1]{\hbox{\boldmath$#1$\unboldmath}}

\begin{document}
%\begin{flushleft}
%LYCEN-2010- \\
%CERN-PH-TH/2010-046\\
%\end{flushleft}

\title{Neutrino and antineutrino quasielastic interactions with nuclei}

\author {M. Martini}
\affiliation{CEA/DAM/DIF, 91297 Arpajon, France}
\author {M. Ericson} 
\affiliation{Universit\'e de Lyon, Univ.  Lyon 1, 
 CNRS/IN2P3, IPN Lyon, F-69622 Villeurbanne Cedex, France}
\affiliation{Theory Group, Physics Department, 
CERN, CH-1211 Geneva, Switzerland}
\author{G. Chanfray}
\affiliation{Universit\'e de Lyon, Univ.  Lyon 1, 
 CNRS/IN2P3, IPN Lyon, F-69622 Villeurbanne Cedex, France}
\author{J. Marteau}
\affiliation{Universit\'e de Lyon, Univ.  Lyon 1, 
 CNRS/IN2P3, IPN Lyon, F-69622 Villeurbanne Cedex, France}

\begin{abstract}
 We investigate the interaction of neutrinos and antineutrinos with nuclei. We explore in particular the role played by the multinucleon  excitations which 
can contaminate the quasielastic cross section. 
For neutrinos the multinucleon term  produces a sizable increase of the quasielastic cross
 section. Part of the effect arises from tensor correlations. For antineutrinos this influence  is smaller
 owing to the axial-vector interference which increases the relative importance of the terms which are
 not affected by these multinucleon excitations.

\end{abstract}

\maketitle

The progresses of neutrino experimental physics have allowed the measurement of several partial cross sections 
in the interaction with nuclei. Most data concern ratios of cross sections 
\cite{Nakayama:2004dp,Hasegawa:2005td,AguilarArevalo:2008xs,:2008eaa,Hiraide:2008eu,AguilarArevalo:2009eb,Kurimoto:2009wq}
but some absolute 
values are now available\cite{Katori:2009du,AguilarArevalo:2009ww,:2010zc}. 
In particular the MiniBooNE collaboration has measured the neutrinos charged current (CC) quasielastic (QE) cross section 
on $^{12}$C for a neutrino beam with average energy of $788$ MeV \cite{:2010zc}. 
In this work ejected nucleons are not detected and the quasielastic cross section is defined as the one for processes 
in which only a muon is detected in the final state.  However it 
is possible that in the neutrino interaction a pion  produced via the excitation of the
 $\Delta$ resonance escapes detection, for instance because it is reabsorbed in the nucleus, leading to multinucleon emission. 
In this case it simulates a quasielastic process. 
The MiniBooNE analysis of the data corrects for this possibility via a Monte-Carlo evaluation of this process. 
The net effect amounts to a reduction of the observed quasielastic cross section. 
After application of this correction the quasielastic cross section thus defined still displays an anomaly 
as compared to a relativistic Fermi gas prediction. The prediction which is  sensitive to the cut-off mass of the axial form factor 
fits the data provided a modified axial form factor is introduced in the calculation, 
with an increase of the axial cut-off mass from the accepted value  $M_A=1.03$ GeV to the value 
 $M_A= 1.35$ GeV; otherwise the calculated cross section is too small \cite{:2010zc}.

On the theoretical side Martini \textit{et al.} \cite{Martini:2009uj} have drawn the attention 
to the existence 
of additional sources of multinucleon emission which are susceptible to produce an apparent increase 
of the ``quasielastic'' cross section. Their evaluation, although approximate, of this contribution shows 
that it is able to account for this apparent increase. 
It stresses in particular the role played by the NN tensor correlations in this enhancement.
 In clarifying this point we will be naturally led to explore the corresponding effect in the case of antineutrinos. 
We will show that antineutrinos can provide a useful test of the origin of this anomaly, which is the aim of the present work.
 
\begin{figure}
\begin{center}
\includegraphics[width=0.95\columnwidth,clip]{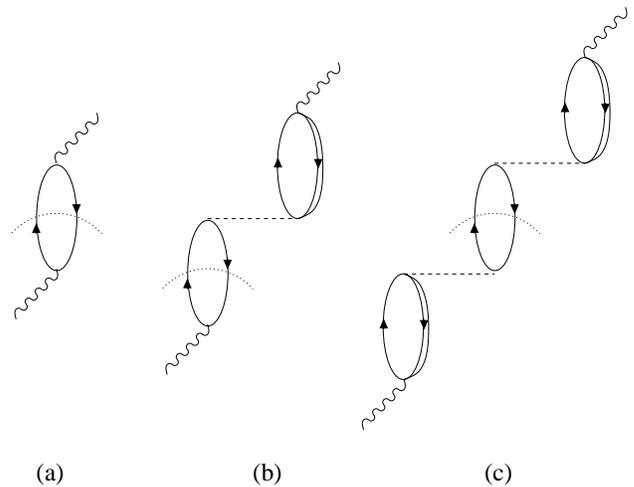}
\caption{Examples of contributions to the quasielastic cross section.
Lowest order contribution from~: (a) $R^{NN}$;  (b) $R^{N\Delta}$ and (c) $R^{\Delta \Delta}$. 
The wiggled lines represent the external probe, the full lines correspond to 
the propagation of a nucleon (or a hole), the double lines to the propagation of a $\Delta$ 
and the dashed lines to an effective interaction between 
nucleons and/or $\Delta$s. The dotted lines show which particles are placed on-shell. } 
\label{N_Delta_1order}
\end{center}
%\vspace*{+0.8cm}
\end{figure}

In our approach, the same as that of  Ref. \cite{Martini:2009uj}, the neutrino cross section on nuclei is expressed 
in terms of the nuclear response functions treated in the random phase approximation (RPA). 
The only nucleon resonance taken into account is the $\Delta$ one. 
Several responses enter this interaction, as exemplified below in a simplified expression 
of the charged current cross section where 
the lepton mass is ignored and the $\Delta$ width is taken to zero. It reads~:
\begin{eqnarray} \label{SIGMANUENU}
\frac{\partial^2\sigma}{\partial\Omega \,\partial k^\prime} & = & 
\frac{G_F^2 \, 
\cos^2\theta_c \, (\gvec{k}^\prime)^2}{2 \, \pi^2} \, \cos^2\frac{\theta}{2} \, 
\left\{ G_E^2 \, (\frac{q_\mu^2}{\gvec{q}^2})^2 \, R_\tau^{NN} \right. \nonumber \\ 
&& +  G_A^2 \, \frac{( M_\Delta - M )^2}{2 \, \gvec{q}^2} \, R_{\sigma\tau (L)}^{N\Delta} 
+  G_A^2 \, \frac{( M_\Delta - M )^2}{\gvec{q}^2} \nonumber \\
&& \times  R_{\sigma\tau (L)}^{\Delta\Delta} 
\!+\!  \left( G_M^2 \frac{\omega^2}{\gvec{q}^2} + G_A^2 \right)\! 
\left( - \frac{q_\mu^2}{\gvec{q}^2} + 2 \tan^2\frac{\theta}{2} \right) \nonumber \\ 
&&\! \times \! \left[R_{\sigma\tau (T)}^{NN}\! +\! 2 R_{\sigma\tau (T)}^{N\Delta}\! 
+\! R_{\sigma\tau (T)}^{\Delta\Delta}\right]\! 
\pm \! 2  G_A  G_M \! \frac{\!k\!+\!k^\prime}{M}  \nonumber \\
&&\! \times  \left. \tan^2\frac{\theta}{2} \, 
\left[R_{\sigma\tau (T)}^{NN} + 2 R_{\sigma\tau (T)}^{N\Delta} 
+ R_{\sigma\tau (T)}^{\Delta\Delta} \right] \right\} 
\end{eqnarray}
where $ G_F $ is the weak coupling constant, $ \theta_c $ the Cabbibo angle, 
$ k $ and $ k^\prime $ the initial and final lepton momenta, $ q_\mu = k_\mu - 
k_{\mu\prime} = ( \omega,\gvec{q} ) $ the four momentum transferred to the 
nucleus, $ \theta $ the scattering angle, $ M_\Delta $ ($ M $) the $\Delta$ 
(nucleon) mass. The plus (minus) sign in Eq. (\ref{SIGMANUENU}) stands for the 
neutrino (antineutrino) case. The existence of this axial-vector interference term is crucial in the present work. 
The various responses  are related to the imaginary part of the corresponding full polarization propagators~:
\begin{equation} \label{eq:6}
R (\omega,q) = -\frac{\mathcal{V}}{\pi} \, \mathrm{Im}[ \Pi(\omega,q,q) ].
\end{equation} 
They are related to the inelastic cross section for a given coupling. For instance for the isospin operator~:
\begin{eqnarray}
\label{eq:2}
R_\tau &=& \sum_n \, 
\langle n | \sum_{j=1}^A \, \tau(j) 
e^{ i \, {\gvec {q}}\cdot {\gvec {x}}_j } | 0 \rangle \nonumber \\
&&\times
\langle n | \sum_{k=1}^A \, \tau(k) 
e^{ i \, {\gvec {q}}\cdot {\gvec {x}}_k } | 0 \rangle^*\, \delta (\omega - E_n + E_0) .  
\end{eqnarray}
Similar expressions apply to the other transition operators. 
An analogous expression applies for a transition to a state with a $\Delta$ excitation which enters the spin-isospin responses. 
In this case the operators are~:  
\begin{equation}
O_{\sigma\tau (L)}^\Delta(j) 
= ( \gvec{S}_j \cdot \hat{\bf{q}} ) \, T_j^\pm, \,\,\, 
O_{\sigma\tau (T)}^\Delta(j) =( \gvec{S}_j \times \hat{\bf{q}} )^i  \, T_j^\pm .  
\end{equation}
 In the case of a spin operator the index L or T refers to the direction of the spin, longitudinal or transverse, 
with respect to the momentum {\bf q}. In the expression (\ref{SIGMANUENU}) 
the upper indices, N or $\Delta$, refer to the type of particles, nucleon 
or $\Delta$, excited by the weak current at the two ends of the RPA chain (see Fig.\ref{N_Delta_1order}). 
In the present numerical evaluations we employ, instead of Eq. (\ref{SIGMANUENU}), 
the full expressions given in the Appendix A of our previous publication \cite{Martini:2009uj} 
for which we refer for more details.

 We have treated these responses in the ring approximation of the RPA 
so as to account for collective effects. A limitation of our description is that it does not incorporate 
final state interactions such as for instance the possibility for a real pion produced by the neutrino to be reabsorbed 
in the nucleus leading to multinucleon ejection. This fraction of the produced pions would thus  be counted in the 
``quasielastic'' events. 
These events have been subtracted in the MiniBooNE analysis through their Monte Carlo evaluation. 
Hence our theory can be confronted to their corrected experimental results.

\begin{figure}
\begin{center}
\includegraphics[width=0.95\columnwidth,clip]{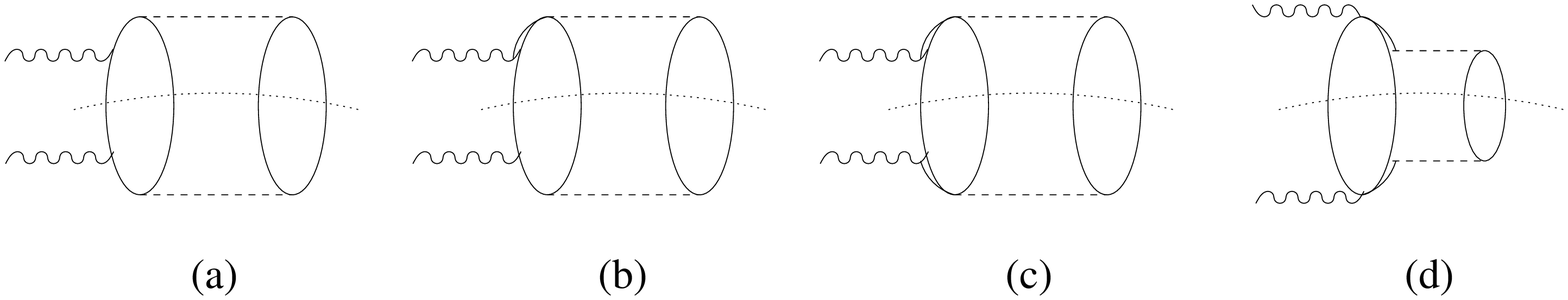}
\caption{Examples of contributions to the two nucleon ejection term.
From~: (a) $R^{NN}$;  (b) $R^{N\Delta}$; (c) and (d) $R^{\Delta \Delta}$. Diagram (d) represents an example of modification of 
the $\Delta$ width in the medium.} 
\label{N_Delta_2p2h}
\end{center}
%\vspace*{+0.8cm}
\end{figure}

To lowest order  the quasielastic cross section is given by the terms in $R^{NN} $, 
whether coming from the isovector interaction, $ R_\tau^{NN}$, or from the isospin
 spin-transverse one, $ R_{\sigma\tau (T)}^{NN}$. 
The isospin spin-longitudinal one is suppressed, for a vanishing lepton mass, 
by a cancellation between the space and time components of the axial current \cite{Martini:2009uj}, 
which is the reason why it is not explicitly written in Eq.(\ref{SIGMANUENU}). 
In the actual calculation its small quasielastic contribution is taken into account. It also contribute to the multinucleon 
ejection term.
In the RPA chain $R^{N\Delta}$ and $R^{\Delta \Delta}$ also contribute to the quasielastic response as illustrated in 
Fig.\ref{N_Delta_1order}. 
The collective effects produce a mild suppression of the quasielastic response due to the repulsive nature of 
the residual interaction. The effect that we want to discuss here is of a different nature. 
It concerns the multinucleon ($np-nh$) ejection in neutrino interactions other than that due to final state interaction. 
Here several sources of  multinucleon emission enter our description and all type the responses $R^{NN}$, $R^{N\Delta}$ 
and $R^{\Delta \Delta}$ contribute. Examples are given in Fig. \ref{N_Delta_2p2h}. 
One of these multinucleon sources  arises from the modification of the $\Delta$ width in the nuclear medium 
(Fig.\ref{N_Delta_2p2h} (d)). 
This effect was introduced and parametrized by Oset and Salcedo \cite{Oset:1987re} in the case of real pion or photon absorption. 
We have used their parametrization  of the modified width, although the kinematics of neutrino interaction 
is different since we are in the space-like region, which could be a source of uncertainty. 
For the other terms not reducible to a modification of a $\Delta$ width of which some examples are shown in 
Fig.\ref{N_Delta_2p2h} we have used a parametrization of Delorme and Guichon \cite{DELGUICH}. They
 exploited a calculation by Shimizu and Faessler \cite{Shimizu:1980kb} 
of the absorptive part of the p-wave pion-nucleus optical potential at threshold,  which 
writes $(4 \pi/2 m_\pi)\,\vec{\nabla}\cdot \textrm{Im}C_0~\rho^2 \vec{ \nabla}$ \cite{Ericson:1988gk}. 
It is known that the absorption mechanism of pions is a two-nucleon one which means 
that in the final state two nucleons are ejected. 
In the many-body language this is a two particle-two hole ($2p-2h$) excitation. 
In Ref. \cite{Shimizu:1980kb} the absorption is described by three types of terms (see Fig.\ref{N_Delta_2p2h}). 
The first one (Fig.\ref{N_Delta_2p2h} (a) with pion lines replacing the weak current ones as in the next diagrams) 
arises from the nucleon-nucleon correlations, essentially from the tensor correlations. 
Another one involves a $\Delta$ excitation (Fig.\ref{N_Delta_2p2h} (c)). 
The third one is an interference between the nucleon correlation and $\Delta$ terms as in Fig.\ref{N_Delta_2p2h} (b).
 In these graphs  the coupling of the pion to the nucleon involves the pion momentum and is of the spin isospin
 type, 
$\vec{\sigma} \cdot \vec{q}  \,\vec{\tau}$, for the nucleon and a similar expression  for $\Delta$ excitation. 
In the optical potential the pion momentum is expressed by the gradient, the remaining part of the optical 
potential with the parameter $\textrm{Im}C_0$ then provides the nuclear $2p-2h$ bare responses to a probe which couples 
to the spin and isospin of the nucleon. 
In principle it is a spin longitudinal coupling but as this term represents a short range effect there is no difference  
in the bare case ({\it i.e.} before the RPA chain) between longitudinal and transverse spin couplings. 
Therefore  it applies to  the bare magnetic part of the vector current and to both the spin-transverse and 
spin-longitudinal parts of the axial current. 
The pion absorption calculation of Ref.\cite{Shimizu:1980kb} is performed  for threshold pion, 
i.e. for a vanishing three-momentum and an energy $\omega=m_{\pi}$, which does not correspond to the neutrino situation.
Delorme \textit{et al.} \cite{DELGUICH} have then introduced in each absorption 
graph the corresponding energy dependence to obtain the  bare $2p-2h$ responses to be inserted in the RPA chain 
needed for neutrino interaction. 
However they have completely ignored the momentum dependence.  
They have left apart the graph (d) of Fig.\ref{N_Delta_2p2h}  since it corresponds to a modification of 
the $\Delta$ width which is taken into account separately through the parametrization of Oset \textit{et al.} \cite{Oset:1987re}. 
In a first approach we have used the  procedure of Delorme \textit{et al.} as such in order 
to evaluate the  bare $2p-2h$ components to be inserted in the RPA chain for the  evaluation of the neutrino cross section. 
In our resulting $2p-2h$ cross section the modification of the $\Delta$ width is not a dominant effect; 
it adds a small $2p-2h$ component (plus a $3p-3h$ one). It turned out that our  overall multinucleon contribution, when
 added to the genuine quasielastic cross section, was able to account for the anomaly without modification of the axial cut-off 
 mass as already reported in Ref.\cite{Martini:2009uj}.
%%%%%%%%%%%%%%%%%

\begin{figure}
\begin{center}
\includegraphics[width=0.95\columnwidth,clip]{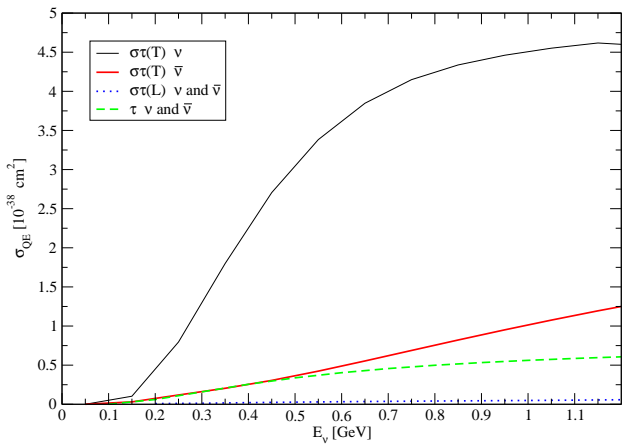}
\caption{Various response contributions to the $\nu_\mu$ and $\bar{\nu}_\mu$-$^{12}$C charged current 
genuine quasielastic cross section. 
Isovector ($\tau$) and isospin spin-longitudinal ($\sigma \tau$ (L)) components are identical 
for $\nu_\mu$ and $\bar{\nu}_\mu$.}
\label{fig_qe}
\end{center}
%\vspace*{+0.8cm}
\end{figure}
Since our parametrization of the $2p-2h$ piece from the extrapolation of pion absorption is questionable, 
as it ignores in particular any momentum dependence, we have also investigated the effect with 
our second parametrization \cite{Martini:2009uj} of the $2p-2h$ contribution beyond the 
one which is reducible to a modification of the Delta width. 
In the second approach we have used a microscopic calculation of Alberico \textit{et al.}
 \cite{Alberico:1983zg} 
specifically aimed at the evaluation of the $2p-2h$ contribution to the isospin spin-transverse response,
measured  
in inclusive $(e,e')$ scattering. Their basic graphs are similar to those of Shimizu and Faessler. In principle this way of evaluation is definitely more satisfactory since the kinematical variables are correctly incorporated
but the results of Ref.\cite{Alberico:1983zg} are available only for a limited set of energy and momenta. 
We have extended this range to cover the neutrino one through an approximate extrapolation. 
We refer to \cite{Martini:2009uj} for the details. 
The result was that the distribution in the energy transfer $\omega$ of the differential neutrino cross section 
is largely modified with a more realistic distribution but once integrated in energy the difference is small. 
The corresponding neutrino and antineutrino total ``quasielastic'' cross sections are practically unchanged.
 
 Now from the way in which this $2p-2h$ contribution is built, the corresponding coupling of the weak 
current to the nucleon or $\Delta$ is a spin isospin one. It is then clear that this $2p-2h$ 
term only affects the magnetic and axial responses which enter the neutrino cross section. 
In the expression (\ref{SIGMANUENU}) these are the terms in $G_A^2, G_M^2$ and the interference term in $ G_AG_M$. 
The isovector response (term in $R_{\tau}$) instead is not affected. 
This difference is the basis for the test that we propose to help elucidate the origin of the anomaly.
The principle is simple. For antineutrinos the interference term in $G_MG_A$ produces a suppression of 
the spin isospin response contribution, while it enhances it for neutrinos. 
In the neutrino case the contribution from the spin isospin terms largely dominates the isovector one.
If the suppression of the spin isospin part for antineutrinos by the interference term modifies the balance in such 
a way that the role of the isovector response in the cross section becomes appreciable, 
the relative role of the $2p-2h$ part will be smaller for antineutrinos. We will show below that this is indeed the case. 
\begin{figure}
\begin{center}
\includegraphics[width=0.95\columnwidth,clip]{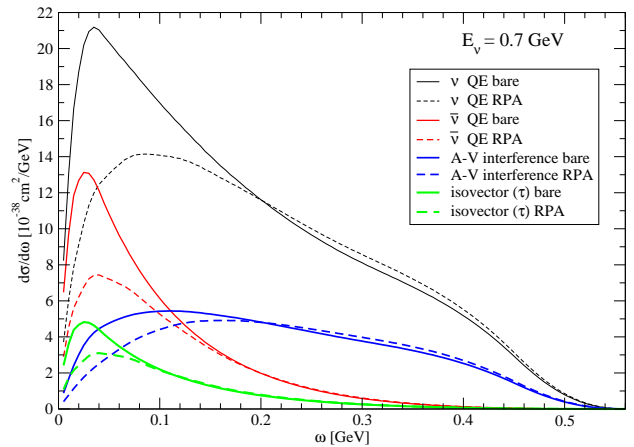}
\caption{Differential CC $\nu_\mu$ and $\bar{\nu}_\mu$ -$^{12}$C cross sections versus the energy transfer.}
\label{fig_diff}
\end{center}
%\vspace*{+0.8cm}
\end{figure}
Figure \ref{fig_qe} displays the various components to the genuine quasielastic (single nucleon ejection) 
cross section on $^{12}$C as a function of the energy both for muon neutrinos and antineutrinos. 
The isovector part is identical in the two cases, as well as the isospin spin-longitudinal one. 
The isospin spin-transverse one instead changes considerably between neutrinos and antineutrinos due 
to the interference term with an appreciable reduction in the antineutrino case. 
As a consequence the isovector relative contribution to the total cross section becomes 
quite significant for antineutrinos which opens the possibility of an experimental test. 
Such details are visible in Fig. \ref{fig_diff} where the differential cross section $\frac{d \sigma}{d \omega}$ 
at $E_{\nu}$=700 MeV is shown in the bare and RPA case, 
both for neutrinos and antineutrinos. In the same figure the axial-vector interference term 
is displayed as well as the isovector component illustrating the importance of the latter in the antineutrino case. 
A remark on the collective nature is in order at this stage. 
One notices the suppression produced by the RPA which follows from the repulsive character of the residual interaction. It 
 is more pronounced at small energies. 
The interference term which spreads over all the energy range is therefore less affected by RPA. 
The antineutrino cross section which peaks at low energies is instead very sensitive to the suppression by the 
collective RPA effects. 
Once integrated over the energy transfer the reduction at a typical energy $E_\nu$=700 MeV 
is somewhat larger for antineutrinos than for neutrinos. A comparison between the two is affected by this difference. 
This could lead to a source of uncertainty in the comparison as the RPA effects which depend on the residual interaction 
have not been tested in this momentum regime. 
However we will show below that it does not prevent the comparison to be significant for our purpose.

\begin{figure}
\begin{center}
\includegraphics[width=0.95\columnwidth,clip]{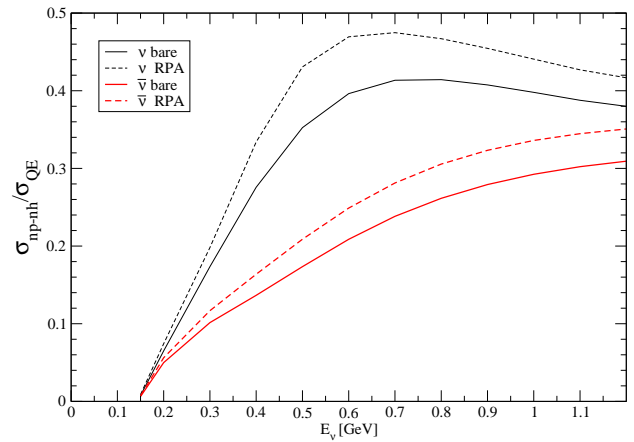}
\caption{Ratio of multinucleon component of ``quasielastic'' cross section on $^{12}$C to the single nucleon one for 
$\nu_\mu$ and $\bar{\nu}_\mu$ as a function of neutrino energy.}
\label{fig_ratio}
\end{center}
%\vspace*{+0.8cm}
\end{figure}

We now turn to the generalized ``quasielastic'' cross section which 
includes the multinucleon contribution. 
\begin{figure}
\begin{center}
\includegraphics[width=0.95\columnwidth,clip]{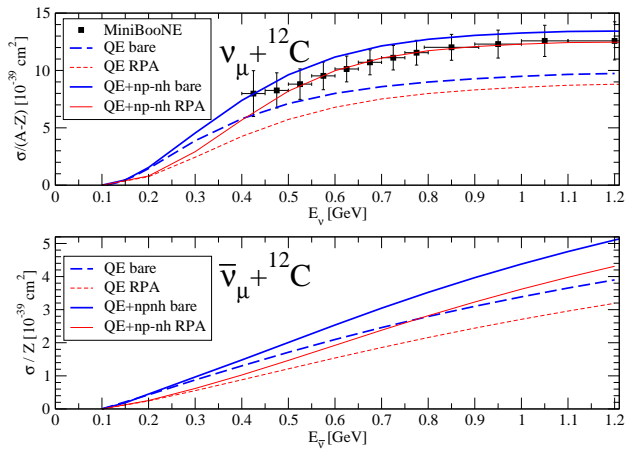}
\caption{$\nu_\mu$ (upper panel) and $\bar{\nu}_\mu$ (lower panel) -$^{12}$C CC ``quasielastic'' cross sections 
per neutron and per proton with and without the multinucleon component as a function of neutrino energy. 
The experimental points are taken from \cite{:2010zc}.}
\label{fig_tot_anu}
\end{center}
%\vspace*{+0.8cm}
\end{figure}
The neutrino and antineutrino genuine and generalized  ``quasielastic'' cross section are plotted in Fig.\ref{fig_tot_anu} 
both in the bare and RPA case. 
As was already discussed in Ref.\cite{Martini:2009uj} the agreement with 
the MiniBooNE experimental neutrino data \cite{Katori:2009du,:2010zc} is better when the $np-nh$ component is added 
to the genuine QE cross section, whether in the free or in the RPA case. 
Our prediction for the generalized neutrino
 ``quasielastic'' cross section shows only a moderate sensitivity to the collective aspects. 
For antineutrinos instead the sensitivity to RPA is somewhat larger but it does not hide the important point that  
the relative importance of the $2p-2h$ term is smaller for antineutrino. This is illustrated  in Fig.\ref{fig_ratio}
which shows the ratio of the multinucleon component to the single nucleon one with and without RPA. In both cases, 
we find that the ratio for antineutrino is reduced as compared to the neutrino one by a factor 1.7 at $E_\nu$=700 MeV. 
In order to eliminate the uncertainties related to the neutrino energy reconstruction, 
we give in Table 1 values of  quasielastic and multinucleon cross sections on $^{12}$C, 
as well as their sum, averaged over the respective 
neutrino \cite{:2010zc} and antineutrino \cite{AguilarArevalo:2008yp} 
MiniBooNE fluxes, so as to provide quantities more in touch with an experimental 
analysis. We gives these quantities both in RPA and in the free case and various situations are possible. 
For instance, if our RPA description holds, the ratio of the generalized ``quasielastic'' 
cross section, which is the measured cross section, to a theoretical free Fermi gas model is 1.22 
for neutrinos and 0.99 for antineutrinos, significantly lower. 
If the extreme case where RPA effects are totally absent the corresponding ratios are 1.37 
for neutrinos and 1.25 for antineutrinos. In all cases the antineutrino numbers are smaller 
and the difference may be detectable, which offers an 
experimental test.
For neutrinos the fit to the quasielastic data in a relativistic Fermi gas description  
required an appreciable increase of the axial cut-off mass \cite{:2010zc}. For antineutrinos 
the increase needed to account for the data in the same model should be smaller since the relative role 
of multinucleon ejection is reduced. This difference offers a possible way to shed light 
on the origin of the anomaly. 
Of course the difference which occurs owing to the fact that the target is not pure carbon but 
CH$_2$ affects exclusively antineutrino cross section reducing trivially the importance of nuclear effects. It has to be taken into account to reach a significant comparison between neutrino and antineutrino cross sections.

\begin{table}[t]
    \begin{center}
        \begin{tabular}{c|ccc|ccc}
            \hline
            $ $       & $ $ & $\nu $ &  $ $     & $ $ & $\bar{\nu}$ & $ $  \\
            $ $       & QE  & np-nh  & QE+np-nh & QE  & np-nh       & QE+np-nh\\
            \hline
            bare $  $ & 7.46&  2.77  &   10.23  & 2.09 & 0.52        &  2.61       \\
            \hline
            RPA   $  $& 6.40&  2.73  &   9.13   & 1.60 & 0.47        &  2.07       \\ 
            \hline
        \end{tabular}
    \caption{    
    \label{table}
    MiniBooNE flux-integrated CC $\nu_\mu$-$^{12}$C and $\bar{\nu}_\mu$-$^{12}$C total cross sections per neutron and per proton respectively 
in unit of 10$^{-39}$ cm$^2$. The experimental CCQE $\nu_\mu$-$^{12}$C value measured by MiniBooNE is 9.429 $\times$ 10$^{-39}$ cm$^2$ with a total normalization error 
of 10.7 \% \cite{:2010zc}.}
    \end{center}
\end{table}

Finally we would like to comment on the absence of final state interactions in our evaluation. 
In addition to the absorption of a produced pion it also ignores the possibility 
for an ejected nucleon to interact with the nucleus emitting another nucleon which leads to a final state 
with two nucleon ejected, the same type of final state as was discussed in this work. 
For the final sate interaction effect all responses are concerned and not only the spin isospin ones. 
This has been taken into account by Benhar and Meloni \cite{Benhar:2009wi} using a spectral function which describes 
the single particle dynamics in an interacting system, deduced from experimental
$ (e,e'p)$ data. Their conclusion is that this inclusion  cannot explain the enhancement of the measured 
neutrino quasielastic cross section. Although the final sates are the same, ($2p-2h$), we are dealing in the present work to
a different type of correlations. These are ground state correlations, 
mostly tensor ones which affect only the spin isospin responses, producing an enhancement of 
the spin isospin sum rule which is reflected in an increase of the ``quasielastic'' cross section. 
The rest is due 
to the $\Delta$ excitation or the interference between the two. These also obviously belong exclusively to the spin isospin sector, 
which is the basis of our test through a comparison between neutrinos and antineutrinos.    
A fully realistic calculation should include both the spectral function effect and the ground state correlations or $\Delta$ ones.     

In summary we have studied the quasielastic neutrino and antineutrino cross section 
in the case of the MiniBooNE experiment where multinucleon ejection is not 
distinguishable from single nucleon production. 
The ``quasielastic'' cross section thus defined  contains a certain proportion of $2p-2h$ and $3p-3h$
 excitations. 
This proportion is large for neutrinos, which may be the interpretation of the increase in 
the axial cut-off mass needed to describe the data in the relativistic Fermi gas. 
For antineutrinos we predict a smaller role of the $2p-2h$ component. 
The reason is that the vector-axial interference term produces a suppression of the spin isospin response
 contribution 
to the cross section leaving a larger role for the isovector response which is not affected be the $2p-2h$ component. 
An experimental confirmation of this difference would signal the fact
that the excess cross section belongs to the spin-isospin channel, thus displaying
the role played by the tensor correlations
in neutrino nucleus interactions.

The antineutrino mode which is actively investigated for the general problems of neutrino oscillations and 
CP violation is also of great relevance for the understanding of the neutrino nucleus interactions.
\vskip 0.2 true cm
We thank D. Davesne, T. Ericson, G.T. Garvey and T. Katori for useful discussions.


\begin{thebibliography}{99}  

%%%%%%%%%%%%%%%%%%%%%%%%%          EXP %%%%%%%%%%%%%%%%%%%%%%%%%%%


%\cite{Nakayama:2004dp}
\bibitem{Nakayama:2004dp}
  S.~Nakayama {\it et al.}  [K2K Collaboration],
  %``Measurement of single pi0 production in neutral current neutrino
  %interactions with water by a 1.3-GeV wide band muon neutrino beam,''
  Phys.\ Lett.\  B {\bf 619}, 255 (2005).
 % [arXiv:hep-ex/0408134].
  %%CITATION = PHLTA,B619,255;%%


%\cite{Hasegawa:2005td}
\bibitem{Hasegawa:2005td}
  M.~Hasegawa {\it et al.}  [K2K Collaboration],
  %``Search for coherent charged pion production in neutrino carbon
  %interactions,''
  Phys.\ Rev.\ Lett.\  {\bf 95} 252301 (2005).
%  [arXiv:hep-ex/0506008].
  %%CITATION = PRLTA,95,252301;%%



%\cite{AguilarArevalo:2008xs}
\bibitem{AguilarArevalo:2008xs}
  A.~A.~Aguilar-Arevalo {\it et al.}  [MiniBooNE Collaboration],
  %``First Observation of Coherent $\pi^0$ Production in Neutrino Nucleus
  %Interactions with $E_{\nu}<$ 2 GeV,''
  Phys.\ Lett.\  B {\bf 664}, 41 (2008).
%  [arXiv:0803.3423 [hep-ex]].
  %%CITATION = PHLTA,B664,41;%%


%\cite{:2008eaa}
\bibitem{:2008eaa}
  A.~Rodriguez {\it et al.}  [K2K Collaboration],
  %``Measurement of single charged pion production in the charged-current
  %interactions of neutrinos in a 1.3 GeV wide band beam,''
  Phys.\ Rev.\  D {\bf 78}, 032003 (2008).
%  [arXiv:0805.0186 [hep-ex]].
  %%CITATION = PHRVA,D78,032003;%%


%\cite{Hiraide:2008eu}
\bibitem{Hiraide:2008eu}
  K.~Hiraide {\it et al.}  [SciBooNE Collaboration],
  %``Search for Charged Current Coherent Pion Production on Carbon in a Few-GeV
  %Neutrino Beam,''
  Phys.\ Rev.\  D {\bf 78} 112004 (2008).
%  [arXiv:0811.0369 [hep-ex]].
  %%CITATION = PHRVA,D78,112004;%%



%\cite{AguilarArevalo:2009eb}
\bibitem{AguilarArevalo:2009eb}
  A.~A.~Aguilar-Arevalo {\it et al.}  [MiniBooNE Collaboration],
  %``Measurement of the \nu_\mu charged current \pi^+ to quasi-elastic cross
  %section ratio on mineral oil in a 0.8 GeV neutrino beam,''
  Phys.\ Rev.\ Lett.\  {\bf 103}, 081801 (2009).
%  arXiv:0904.3159 [hep-ex].
  %%CITATION = ARXIV:0904.3159;%%

%\cite{Kurimoto:2009wq}
\bibitem{Kurimoto:2009wq}
  Y.~Kurimoto {\it et al.}  [SciBooNE Collaboration],
  %``Measurement of Inclusive Neutral Current Neutral Pion Production on Carbon
  %in a Few-GeV Neutrino Beam,''
  Phys.\ Rev.\  D {\bf 81}, 033004 (2010).
%  arXiv:0910.5768 [hep-ex].
  %%CITATION = ARXIV:0910.5768;%%

%\cite{Katori:2009du}
\bibitem{Katori:2009du}
  T.~Katori  [MiniBooNE Collaboration],
  %``First Measurement of Muon Neutrino Charged Current Quasielastic (CCQE)
  %Double Differential Cross Section,''
  AIP Conf.\ Proc.\  {\bf 1189}, 139 (2009).
%  arXiv:0909.1996 [hep-ex].
  %%CITATION = ARXIV:0909.1996;%%

%\cite{AguilarArevalo:2009ww}
\bibitem{AguilarArevalo:2009ww}
  A.~A.~Aguilar-Arevalo {\it et al.}  [MiniBooNE Collaboration],
  %``Measurement of $\nu_\mu$ and $\bar{\nu}_\mu$ induced neutral current single
  %$\pi^0$ production cross sections on mineral oil at $E_\nu\sim O(1 GeV)$,''
  Phys.\ Rev.\  D {\bf 81}, 013005 (2010).
%  [arXiv:0911.2063 [hep-ex]].
  %%CITATION = PHRVA,D81,013005;%%

%\cite{:2010zc}
\bibitem{:2010zc}
  A.~A.~Aguilar-Arevalo {\it et al.}  [MiniBooNE Collaboration],
  %``First Measurement of the Muon Neutrino Charged Current Quasielastic Double
  %Differential Cross Section,''
  arXiv:1002.2680 [hep-ex].
  %%CITATION = ARXIV:1002.2680;%%


%

%%%%%%%%%%%%%%%%%%%%%%%%%%%%%%%%%%%%%%%%%%%%%%%%%%%%%%%%%%

%\cite{Martini:2009uj}
\bibitem{Martini:2009uj}
  M.~Martini, M.~Ericson, G.~Chanfray and J.~Marteau,
  %``A unified approach for nucleon knock-out, coherent and incoherent pion
  %production in neutrino interactions with nuclei,''
  Phys.\ Rev.\  C {\bf 80}, 065501 (2009).
%  [arXiv:0910.2622 [nucl-th]].
  %%CITATION = PHRVA,C80,065501;%%

\bibitem{Oset:1987re} E.~Oset and L.~L.~Salcedo, Nucl.\ Phys.\  A {\bf 468} 631 (1987).


\bibitem{DELGUICH} J. Delorme, P.A.M. Guichon, in 
\textit{Proceedings of $10^e$ biennale de physique nucl\'eaire, Aussois, France, March 6-10, 1989}, LYCEN report 8906, p. C.4.1. 
%,also in the \textit{Proceedings of the 5th french-japanese symposium on nuclear physics, Dogashima, Izu, September 26-30, 1989}, 
%edited by K. Shimizu and O. Hashimoto, p.66 (University of Tokio Press, Tokio, 1990).
\bibitem{Shimizu:1980kb} K.~Shimizu and A.~Faessler, Nucl.\ Phys.\  A {\bf 333} 495 (1980).

%\cite{Ericson:1988gk}
\bibitem{Ericson:1988gk}
  T.~E.~O.~Ericson and W.~Weise,
\textit{Pions and Nuclei, The Interrnational Series of Monographs on Physics} 
(Clarendon, Oxford, UK, 1988), Vol. 74. 


%\cite{Alberico:1983zg}
\bibitem{Alberico:1983zg}
  W.~M.~Alberico, M.~Ericson and A.~Molinari,
  %``The Role Of Two Particles - Two Holes Excitations In The Spin - Isospin
  %Nuclear Response,''
  Annals Phys.\  {\bf 154}, 356 (1984).
  %%CITATION = APNYA,154,356;%%


%\cite{AguilarArevalo:2008yp}
\bibitem{AguilarArevalo:2008yp}
  A.~A.~Aguilar-Arevalo {\it et al.}  [MiniBooNE Collaboration],
  %``The Neutrino Flux prediction at MiniBooNE,''
  Phys.\ Rev.\  D {\bf 79}, 072002 (2009).
%  [arXiv:0806.1449 [hep-ex]].
  %%CITATION = PHRVA,D79,072002;%%


%\cite{Benhar:2009wi}
\bibitem{Benhar:2009wi}
  O.~Benhar and D.~Meloni,
  %``Impact of nuclear effects on the determination of the nucleon axial mass,''
  Phys.\ Rev.\  D {\bf 80}, 073003 (2009).
%  [arXiv:0903.2329 [hep-ph]].
  %%CITATION = PHRVA,D80,073003;%%


\end{thebibliography}
\end{document}